\documentclass[prl,twocolumn,superscriptaddress,longbibliography,preprintnumbers,floatfix,nofootinbib]{revtex4-1}

\usepackage{url}
\usepackage{xspace}
\usepackage{dsfont}
\usepackage{amssymb}
\usepackage{amsmath}
\usepackage{graphicx}
\usepackage[caption=false]{subfig}
\usepackage[colorlinks=true,citecolor=blue]{hyperref}
\usepackage{multirow}
\usepackage{hhline}

\begin{document} 

\title{Incorporating Physical Priors into Weakly-Supervised Anomaly Detection}

\author{Chi Lung Cheng}
\email{ccheng84@wisc.edu}
\affiliation{Department of Physics, University of Wisconsin, Madison, WI 53706, USA}
\affiliation{Physics Division, Lawrence Berkeley National Laboratory, Berkeley, CA 94720, USA}

\author{Gup Singh}
\email{gupsingh@berkeley.edu}
\affiliation{Physics Division, Lawrence Berkeley National Laboratory, Berkeley, CA 94720, USA}

\author{Benjamin Nachman}
\email{bpnachman@lbl.gov}
\affiliation{Physics Division, Lawrence Berkeley National Laboratory, Berkeley, CA 94720, USA}
\affiliation{Berkeley Institute for Data Science, University of California, Berkeley, CA 94720, USA}

\begin{abstract}

We propose a new machine-learning-based anomaly detection strategy for comparing data with a background-only reference (a form of weak supervision).  The sensitivity of previous strategies degrades significantly when the signal is too rare or there are many unhelpful features.  Our Prior-Assisted Weak Supervision (PAWS) method incorporates information from a class of signal models to significantly enhance the search sensitivity of weakly supervised approaches.  As long as the true signal is in the pre-specified class, PAWS matches the sensitivity of a dedicated, fully supervised method without specifying the exact parameters ahead of time. On the benchmark LHC Olympics anomaly detection dataset, our mix of semi-supervised and weakly supervised learning is able to extend the sensitivity over previous methods by a factor of 10 in cross section.  Furthermore, if we add irrelevant (noise) dimensions to the inputs, classical methods degrade by another factor of 10 in cross section while PAWS remains insensitive to noise.  This new approach could be applied in a number of scenarios and pushes the frontier of sensitivity between completely model-agnostic approaches and fully model-specific searches.

\end{abstract}

\maketitle
\flushbottom

\section{Introduction}
\label{sec:intro}

Even if we could enumerate all potential new physics scenarios discoverable at particle physics experiments, it would be impossible to conduct a search for each one.  Therefore, physics model agnostic \textit{anomaly detection} searches are an essential complement to the traditional physics model specific searches.  Modern machine learning has ignited a rapidly growing literature on innovative anomaly detection strategies that are projected to qualitatively extend the sensitivity of the particle physics search program~\cite{Kasieczka:2021xcg,Aarrestad:2021oeb,Karagiorgi:2022qnh}. Several of these strategies have already been applied to data~\cite{collaboration2020dijet,ATLAS:2023ixc,ATLAS:2023azi,CMS-PAS-EXO-22-026,Shih:2021kbt,Shih:2023jfv,Pettee:2023zra,Sengupta:2024ezl}.

One powerful set of anomaly detection strategies are based on classifiers trained to distinguish data from a background-only reference~\cite{Collins:2018epr,Collins:2019jip,DAgnolo:2018cun,DAgnolo:2019vbw,Andreassen:2020nkr,Nachman:2020lpy,Benkendorfer:2020gek,Hallin:2021wme,dAgnolo:2021aun,Golling:2022nkl,Hallin:2022eoq,Golling:2023yjq,Sengupta:2023xqy}.  If the background only reference is statistically identical to the background in the data, then the classifier will be able to pick up on any non-background component in the data.  This weakly supervised strategy is asymptotically optimal~\cite{Nachman:2020lpy} and various methods are distinguished by how they construct the background reference, how they construct the classifier, and how they determine the statistical compatibility of the data with the reference.

Until recently, most studies of the background reference strategy used low-dimensional, high-level features.  Additional features can degrade the sensitivity of the search, especially when the new features have the same spectrum for signal and background~\cite{Finke:2023ltw,Freytsis:2023cjr}.  The authors of Ref.~\cite{Buhmann:2023acn} have shown how to employ the background reference strategy to high-dimensional, low-level features, but the sensitivity degrades rapidly with low signal cross sections. We hypothesize that existing methods use overly flexible classifiers, which allow them to model unphysical features, thereby reducing sensitivity. Our solution is to incorporate physical priors to reduce the space of functions.  

To this end, we introduce a Prior-Assisted Weak Supervision (PAWS) strategy for anomaly detection.  Our approach starts with specifying a class of signal models, parameterized by $\theta\in\mathbb{R}^N$. These parameters could be masses, couplings, or other physical properties.  As a first step, we use simulated signal samples to pre-train neural network classifiers parameterized by $\theta$~\cite{Cranmer:2015bka,Baldi:2016fzo} using features $x\in\mathbb{R}^M$.  Then, we use the same classifier to distinguish the data from the background-only reference.  In the second classification step, we freeze all of the network weights and promote $\theta$ to be learnable.  Whereas previous methods scale poorly with $M$, PAWS is independent of $M$ and degrades instead with $N$.  In other words, the set of PAWS classifiers exist on a lower-dimensional manifold of functions than a typical unconstrained classifier.  By only considering these physical classifiers, the sensitivity may be significantly improved compared to more general searches without the pretraining step.  Additionally, since $\theta$ is optimized like any other neural network network parameter via gradient descent, far fewer parameter values are checked compared to a classical search that would scan over values, which is itself intractable in many dimensions. The tradeoff is that we assume the anomaly is in our signal class (and can be reasonably simulated).  This is a strong constraint, but this still gives us access to large classes of physics models that cannot be searched for individually.

There have been a number of studies that have investigated related aspects of our approach.  References~\cite{Park:2020pak,Khosa:2020qrz} use a set of signal models to indirectly improve the performance to out-of-distribution signals.  A pair of recent studies showed that less expressive classifiers (boosted decision trees) can help mitigate degradation due to irrelevant features~\cite{Finke:2023ltw,Freytsis:2023cjr}.  References~\cite{Beauchesne:2023vie,Golling:2024abg,Mikuni:2024qsr} use transfer learning from models trained on similar tasks to inject some physics information through the initialization of the classifier.  Instead of pre-training on signal models, Ref.~\cite{Das:2023bcj} recycles a background prediction to enhance the sensitivity to small signals.  Such an approach is complementary to our proposal.  Another way of viewing our approach is as a variation of neural simulation-based inference~\cite{Cranmer_2020}, where the signal fraction $\mu$ and $\theta$ are inferred from the data.  In the anomaly detection case, we are (mostly) interested in knowing if the classifier learns anything non-trivial, but a byproduct of the approach is that there is also an estimate of the physics parameters.

\section{Dataset}
\label{sec:data}

As a representative example, we use collisions at the Large Hadron Collider (LHC). We build on the LHC Olympics (LHCO) R\&D dataset~\cite{Kasieczka:2021xcg,LHCOlympics} and the extended background dataset from Ref.~\cite{Hallin:2021wme} for our studies.  The original dataset consists of generic quark-gluon scattering processes to produce dijet events as the Standard Model background and $W' \rightarrow X Y $ with $X, Y \rightarrow q\bar{q}$ as signal.  Both processes were simulated using \textsc{Pythia}~8.219~\cite{Sjostrand:2006za, Sjostrand:2014zea} and \textsc{Delphes}~3.4.1~\cite{deFavereau:2013fsa, Mertens:2015kba}.  When $m_{W'}\gg m_X, m_Y$, the $X$ and $Y$ particle decay products will be mostly captured inside a single jet each and their invariant mass will be close to $m_{W'}$.  The original dataset had $m_{W'}$ = 3.5\,TeV, $m_{X}$ = 500\,GeV, and $m_{Y}$ = 100\,GeV.  To train our parameterized classifier and to demonstrate the performance on a variety of signal models, we use the same simulator settings to generate an entire grid with $m_X,m_Y<600$\,GeV in increments of 50\,GeV.  The upper mass bound ensures that the quarks are well-contained within a single jet.  In addition to the two-prong $X,Y\rightarrow qq$ decay, we also include the three-prong $Y\rightarrow qqq$ decay from Ref.~\cite{LHCOlympics} to increase the dimensionality of our parameter space.  In total, each decay features 144 signal models, with around 100k events each.  The parameters of our signal model set are $\theta=\{m_X, m_Y,\alpha\equiv\text{BR}(Y\rightarrow qqq)\}$, so $N=3$. When it would be ambiguous, we will denote $\hat{\theta}$ to mean the true values instead of the fitted values.

The reconstructed particles of each event are clustered into jets using the anti-$k_{T}$ algorithm~\cite{Cacciari:2005hq, Cacciari:2011ma, Cacciari:2008gp} with $R=1.0$.  All events are required to satisfy $p_{T} >$ 1.2\,TeV.  The two highest $p_T$ jets $J_1$ and $J_2$ are sorted such that $m_{J_1} > m_{J_2}$.  Each event is characterized by the same six features ($M=6$) that have been widely used in LHCO studies: $x=\{m_{J_1}, m_{J_2}, \tau_{21}^{J_1}, \tau_{21}^{J_2}, \tau_{32}^{J_1}, \tau_{32}^{J_2}\}$, where the $n$-subjettiness ratios are defined as $\tau_{ij} \equiv \tau_{i} /\tau_{j}$~\cite{Thaler:2010tr,Thaler:2011gf}.  These are sensitive to the two- ($\tau_{21}$) or three- ($\tau_{32}$) prong nature of the jets.  

The signal is resonant in $m_{JJ}$ and a full analysis would scan over $m_{JJ}$ to form signal regions and sideband regions.  As in many previous studies, for simplicity, we focus on the one signal region $m_{JJ} \in [3.3, 3.7]$\,TeV that contains the signal.  The sideband $m \not\in [3.3, 3.7]$\,TeV would be used to both estimate the background-only distribution of features $p_\text{B}(x)$ for training classifiers and to estimate the total number of background events after applying a threshold in the classifier. Previous studies have shown that deep learning is very effective for estimating $p_\text{B}$ from data (e.g. Ref.~\cite{Hallin:2021wme}) and so we assume it is already known; instead, we focus on how to use samples from $p_\text{B}$ to achieve signal sensitivity.  Future experimental studies would synthesize all of these components into a full analysis.  In practice, our reference dataset is statistically identical to and has the same size\footnote{Previous studies have shown that there may be advantages to oversampling the reference~\cite{Hallin:2021wme} - this could be combined with our approach in future studies.} as the background in our actual `data'.


\section{Methods}
\label{sec:method}

Given features $x$ and parameters $\theta$, we start with a fully supervised parameterized classifier $f_\text{FS}(x,\theta)$ trained with the binary cross entropy (BCE) loss:

\begin{align}
\label{eq:bce}
    -\sum_{(x,\theta)\in S} \log(f_\text{FS}(x,\theta)) - \sum_{(x,\theta)\in B\times\Theta} \log(1 - f_\text{FS}(x,\theta))\,,
\end{align}
where $S$ are signal events specified by their features and their parameters, and $B$ are background events specified by their features.  The signal dataset is generated such that every event uses parameters in the set $\Theta$ and then $X_i\sim p_\text{S}(\theta_i)$ for $\theta_i\in\Theta$.  The notation $B\times\Theta$ in Eq.~\ref{eq:bce} represents the fact that the $\theta$ values used for the background sample are independent from the features and distributed in the same way as for the signal.  This means that $\theta$ itself is not useful for classification. 

Asymptotically, the optimal $f_\text{FS}$ from Eq.~\ref{eq:bce} will approximate

\begin{align}
    f_\text{FS}^*&\approx\frac{P_\text{S}(x|\theta)p(\theta)}{P_\text{S}(x|\theta)p(\theta)+P_\text{B}(x)p(\theta)}
    =\frac{P_\text{S}(x|\theta)}{P_\text{S}(x|\theta)+P_\text{B}(x)}\,.
\end{align}
where $P_i$ is the probability density $p_i$ multiplied by the number of events $N_i$ used in training.  Define $\kappa(\theta)\equiv N_B/N_S(\theta)$; the $\theta$ dependence is due to acceptance effects that results in slightly uneven number of signal events in training as a function of the particle masses. We can extract the likelihood ratio from the classifier: $p_\text{S}(x|\theta)/p_\text{B}(x)\approx \kappa(\theta)\,f^*_\text{FS}(x)/(1-f^*_\text{FS}(x))\equiv \Lambda_\text{FS}(x|\theta)$.

Our weakly supervised classifiers are also trained using the BCE loss~\cite{Metodiev:2017vrx,Collins:2018epr,Collins:2019jip}:

\begin{align}
\label{eq:bceWS}
    -\sum_{(x)\in D} \log(f_\text{WS}(x)) - \sum_{(x)\in R} \log(1 - f_\text{WS}(x))\,,
\end{align}
where $D$ is the data with $p_\text{D}(x)=\mu p_S(x|\theta)+(1-\mu)p_B(x)$ and $R$ is the reference with $p_R=p_B$.  Asymptotically, the optimal $f_\text{WS}$ from Eq.~\ref{eq:bceWS} will approximate

\begin{align}
 \label{eq:ws}
    f_\text{WS}^*(x)&\approx\frac{P_D(x)}{P_D(x)+P_R(x)}
    =\frac{\mu \frac{p_S(x|\theta)}{p_B(x)}+(1-\mu)}{\mu \frac{p_S(x|\theta)}{p_B(x)}+2(1-\mu)}\,.
\end{align}

Previous approaches parameterized $f_\text{WS}$ as a generic neural network.  In our new approach PAWS, we insert $f^*_\text{FS}$ by replacing the likelihood ratio in Eq.~\ref{eq:ws} with $\Lambda_\text{FS}$:

\begin{align}
    \label{eq:sws}
    \tilde{f}_\text{WS}(x|\mu,\theta)=\frac{\mu \Lambda_\text{FS}(x|\theta)+(1-\mu)}{\mu \Lambda_\text{FS}(x|\theta)+2(1-\mu)}\,,
\end{align}
with the functional form of $\Lambda_\text{FS}$ fixed and only $\theta$ and $\mu$ are optimized.  When $p(x|\theta)$ factorizes, it is possible to add additional structure. This allows us to consider both two-prong and three-prong decays of the $Y$ particle.  
 In particular, we train separate supervised classifiers for each decay mode and then mix the likelihood ratios according to the branching fraction $\alpha$.

\indent Each model $f_\text{FS}$ and $f_\text{WS}$ is parameterized as a three-layer fully-connected neural network with node configurations of (256, 128, 64). The model $\tilde{f}_\text{WS}$ is parameterized in terms of 
 $f_\text{FS}$ according to Eq.~\ref{eq:sws}.  The networks are implemented in TensorFlow~\cite{tensorflow2015-whitepaper} and optimized using Adam~\cite{Kingma:2014vow} with early stopping.  The data are divided so that 50\% are used for training, 25\% are used for validation and the rest for testing.  In weakly supervised training, 50\% of the background is in $D$, 50\% is in $R$ and the signals are included in $D$ according to $\mu$ and $\alpha$.  In practice, a $k$-fold procedure would be used to ensure that all data could participate in both training and testing~\cite{Collins:2018epr,Collins:2019jip,collaboration2020dijet}.  To improve robustness, we train 20 models with the same input data but different random network initializations.  In the case of $\tilde{f}_\text{WS}$, this corresponds to sampling uniformly in the range of masses with fixed values of $\mu=\exp(-4)$ and $\alpha=\exp(-1)$.  Per event, the scores of the 10 models with the lowest validation loss are averaged. As there is a local minimum when $m_X\leftrightarrow m_Y$, we always check the symmetric point after training and take the model with the lowest validation loss.  
 We also found that it was useful to regularize the network with gradient clipping and the loss function with a penalty for $\theta$ that grows exponentially with the distance away from physically valid regions. That is, we constrain $m_{X}$ and $ m_{Y}$ to lie within the range $\left[50, 600\right]$ GeV, penalizing values outside the boundary.  The $f_\text{FS}(x|\theta)$ model was validated against a series of models trained with a particular value of $\theta$ and excellent agreement was observed.   We did not extensively optimize the hyperparameters; the full set of choices can be found in the public software.  

Our figure of merit for comparing different approaches is the Significance Improvement Characteristic defined as $\text{SIC} = \varepsilon_{S} / \sqrt{\varepsilon_{B}}$, where $\varepsilon_{S}$ and $\varepsilon_{B}$ are the signal and background efficiencies, respectively.  The SIC represents the multiplicative factor by which the signal significance increases for a selection on the network corresponding to $\varepsilon_{B}$.  In practice, we do not know \textit{a priori} what selection to make, so we consider the value of the SIC at a fixed value of $\varepsilon_{B}=0.1\%$, which we can determine using $R$.  This is also more numerically stable than the maximum of the SIC curve which has often been used in the past.  In order to quantify the uncertainty from the finite data statistics, we compute the SIC for 10 random realizations of the training inputs.

\section{Results}
\label{sec:results}

We find that PAWS works well across the entire grid of $m_X$ and $m_Y$ values.  To illustrate the performance in detail, we focus on two benchmark points: $m_{X}$, $m_{Y} = (300, 300)$\,GeV and $(100, 500)$\,GeV, showcasing scenarios of highly symmetric and asymmetric masses, respectively.  The asymmetric model is also the same as the one in the R\&D dataset of the LHCO.  The branching ratio for the three-prong decay of the $Y$ particle is set to $\alpha=0.5$ and we scan over $\mu$ to show our sensitivity to low levels of signal.  

\begin{figure*}[ht!]
    
    \begin{minipage}[b]{0.9\linewidth}
        \centering
        \includegraphics[width=\linewidth]{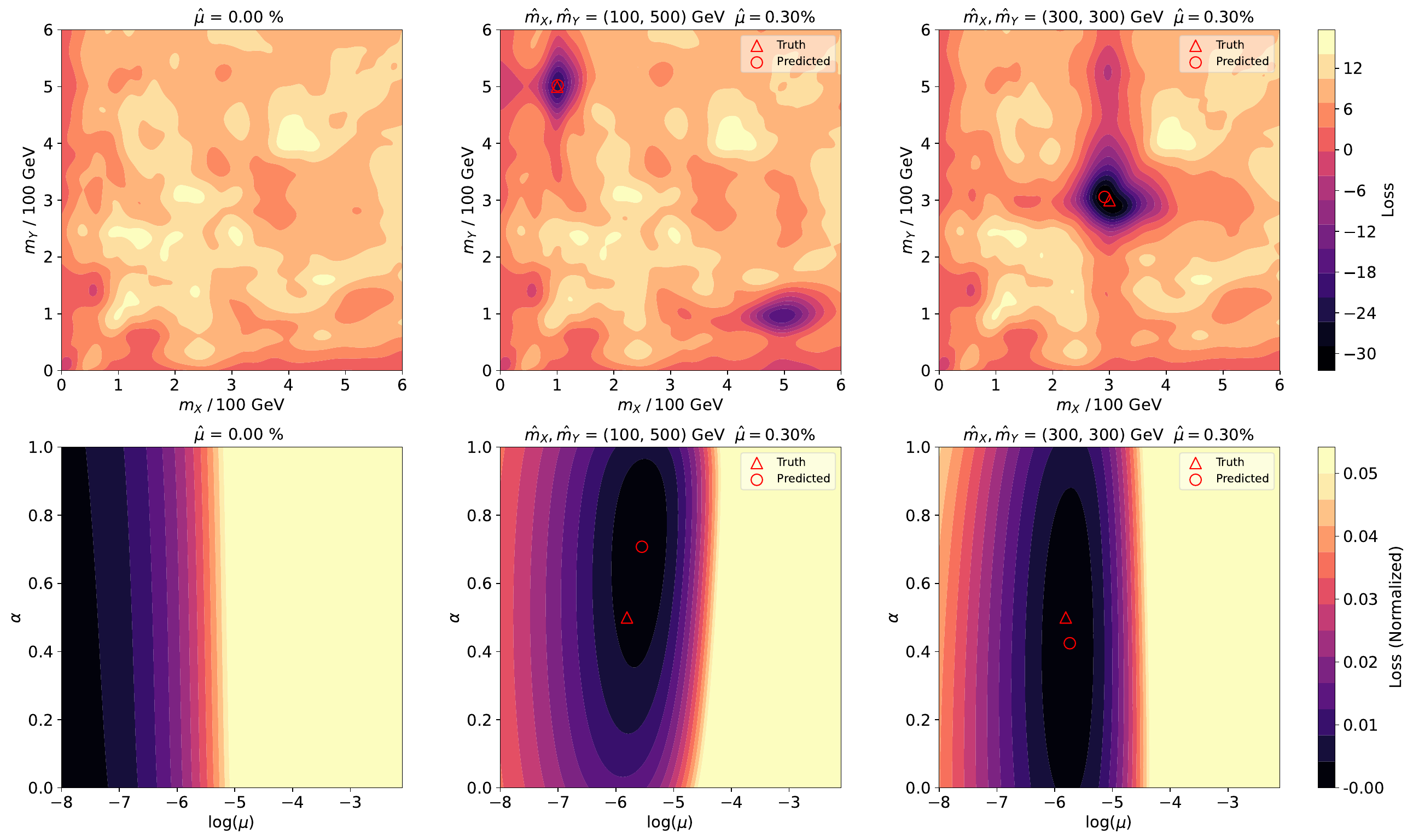}
    \end{minipage}
     \caption{The weakly-supervised loss as a function of $m_X$, $m_Y$, $\mu$ and $\alpha$. The top row varies $m_X$ and $m_Y$ with fixed true values $\mu$ and $\alpha$, while the bottom row scans $\mu$ and $\alpha$ with $m_X$ and $m_Y$ fixed to the true values. The left column shows the loss for a pure background sample, while the middle and right columns correspond to a 0.3\% signal injection at $(\hat{m}_X,\hat{m}_Y) = (100, 500)$\,GeV and $(300, 300)$\,GeV, respectively. Loss values in the $\mu$-$\alpha$ scans are min–max normalized per plot for better visual comparison across different signal injections, which would otherwise have vastly different scales due to the behavior of $\mu$.}
    
    \label{fig:losslandscape}
\end{figure*}

Unlike traditional weak supervision with fully flexible neural networks, for PAWS, the learned parameters are physical and thus directly interpretable when the anomaly is in the pretrained model class.  Figure~\ref{fig:losslandscape} shows the weakly-supervised loss (Eq.~\ref{eq:bceWS}) as a function of $m_X$ and $m_Y$ for $\alpha$ and $\mu$ fixed to their correct values.  The $\mu$ values are chosen to approximately correspond to just before and then just about where the classical weakly supervised approach becomes sensitive.  All four plots in Fig.~\ref{fig:losslandscape} show clear minima at the correct values of $m_X$ and $m_Y$.  In the asymmetric case, there is also a (local) minimum for the incorrect point with $m_X\leftrightarrow m_Y$.  The classifier at this point is still very effective because $m_{J_1}$ and $m_{J_2}$ have nearly the same spectra as for the right mass point.  

\begin{figure*}[ht!]
    \begin{minipage}[b]{0.35\linewidth}
        \centering
        \includegraphics[width=\linewidth]{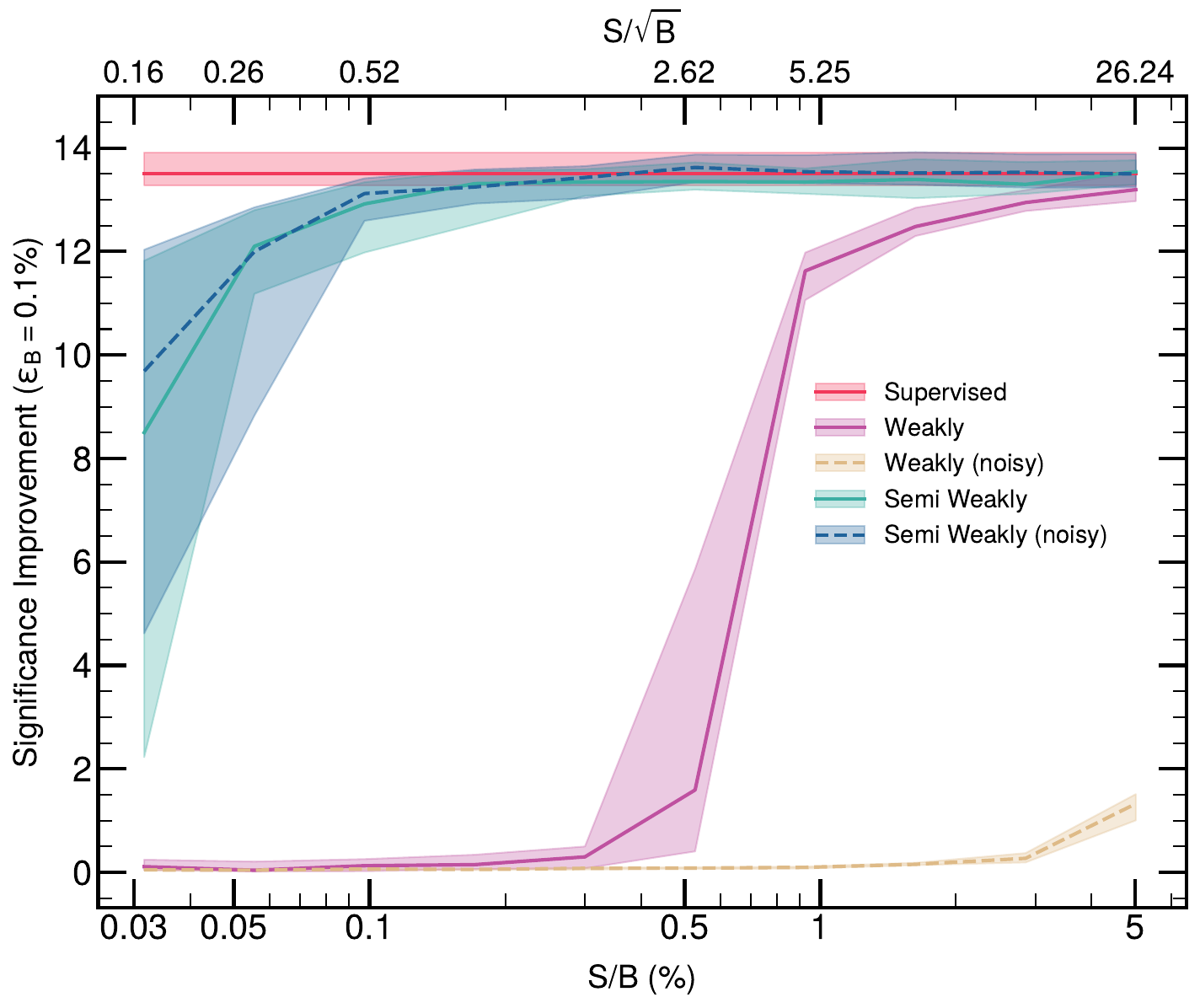}
    \end{minipage}
    \begin{minipage}[b]{0.35\linewidth}
      \centering
        \includegraphics[width=\linewidth]{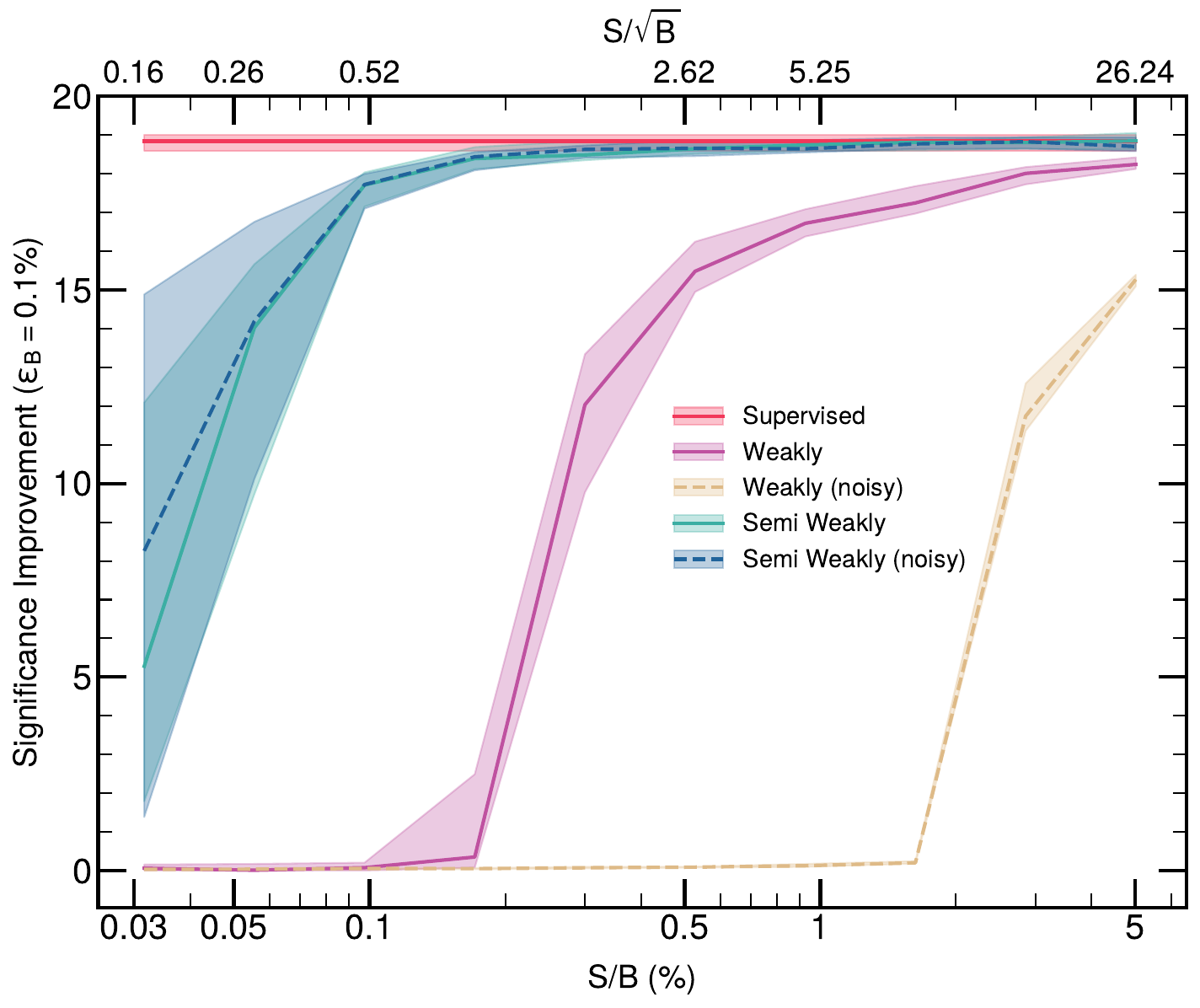}
    \end{minipage}

    \begin{minipage}[b]{0.35\linewidth}
        \centering
        \includegraphics[width=\linewidth]{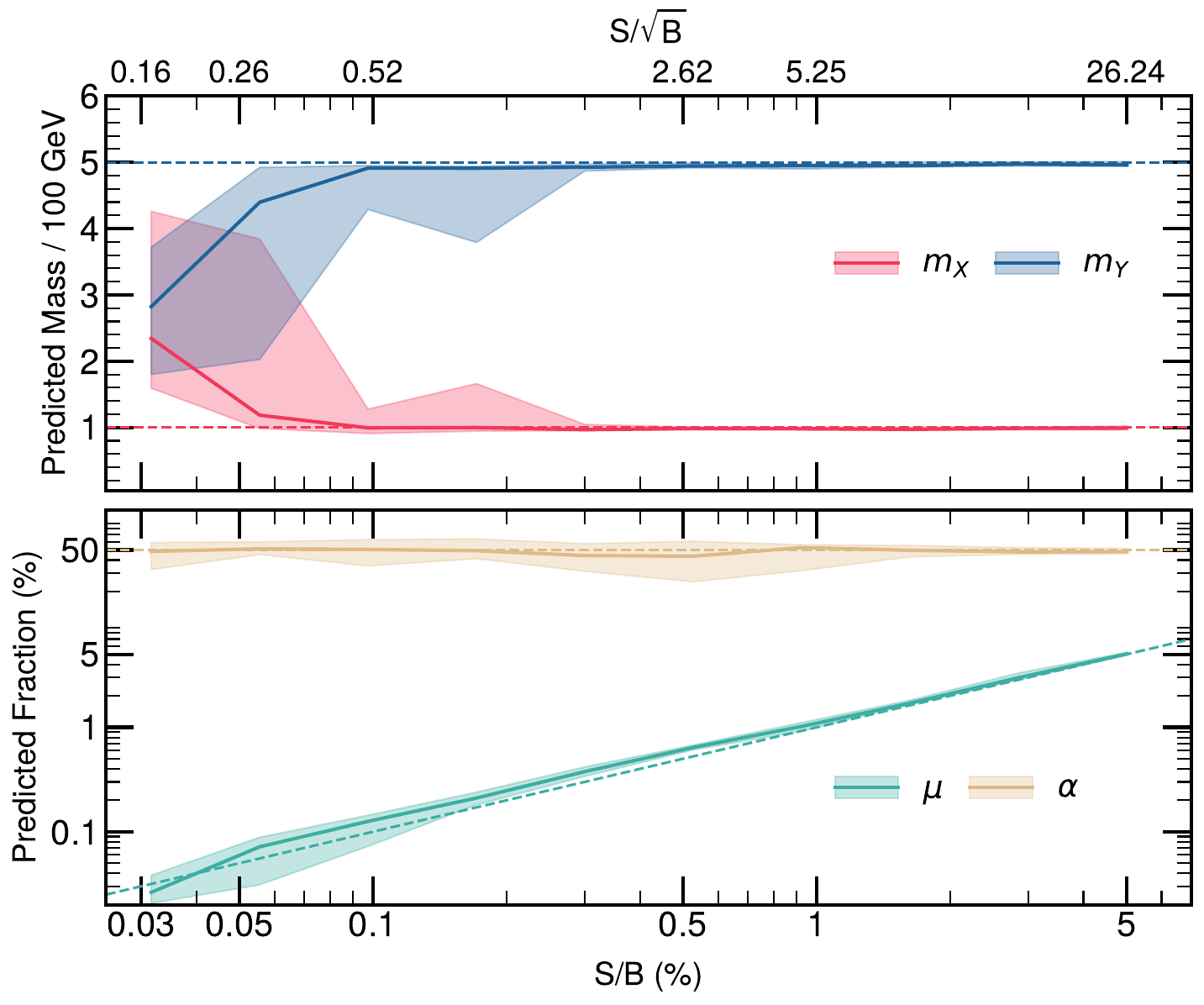}
    \end{minipage}
    \begin{minipage}[b]{0.35\linewidth}
      \centering
        \includegraphics[width=\linewidth]{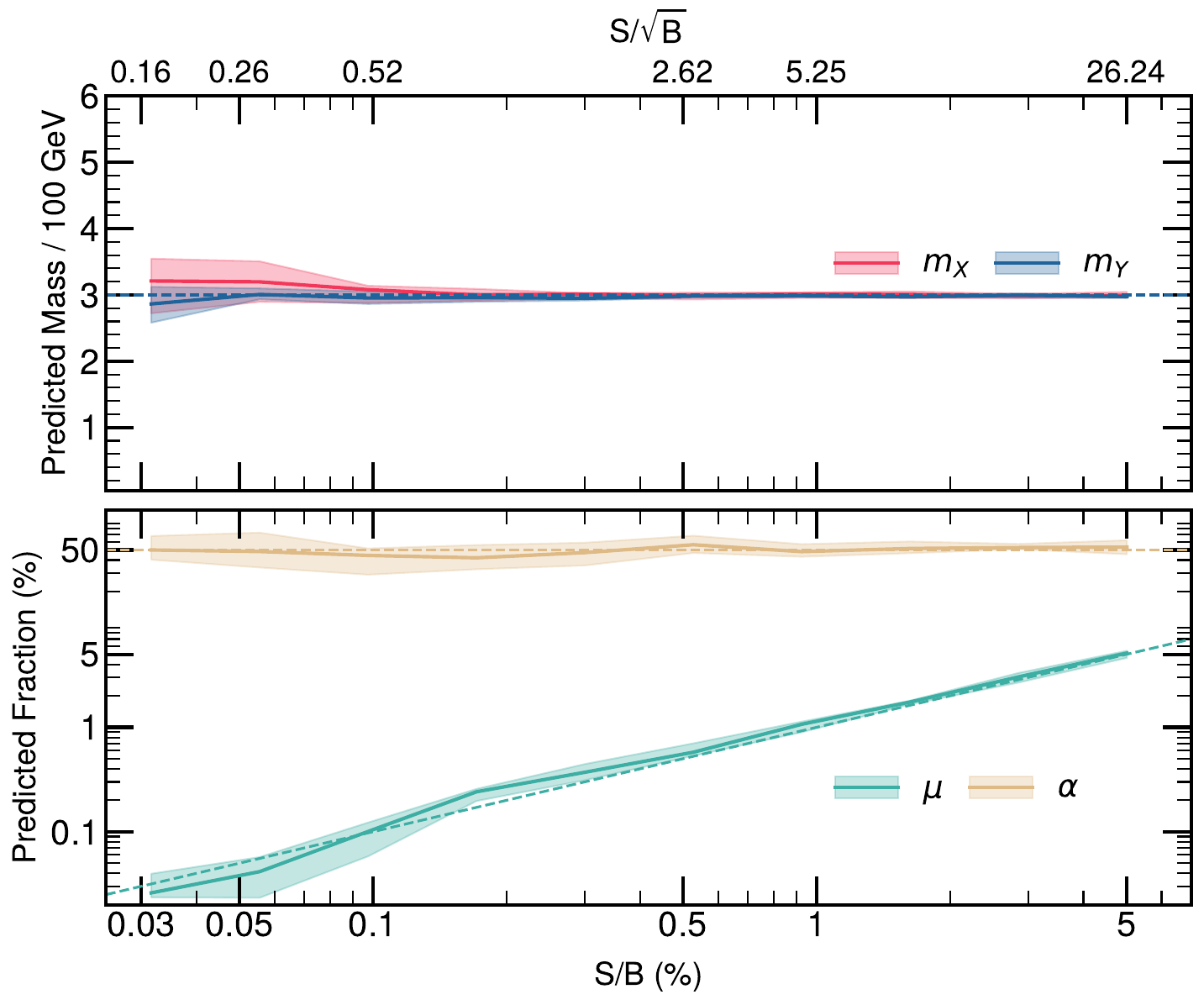}
    \end{minipage}
    
    \caption{Significance improvement at a fixed value of $\varepsilon_{B}=0.1\%$ (top) and predicted parameter values (bottom) with varying amounts of signal injections for $(m_X, m_Y) = (100, 500)$\,GeV (left) and $(300, 300)$\,GeV (right).  The shaded uncertainty bands quantify the standard deviation (around the median) from retraining the respective models with random realizations of the inputs.  The top axis indicates the starting significance of the signal ($S/\sqrt{B}$). The dashed lines from the top plots include scenarios where 10 dimensions of pure noise are added to the inputs.  The dashed lines from the bottom plots indicate the truth values of the parameters represented by the inputs.}
    
    \label{fig:muscan}
\end{figure*}

When the correct mass points are chosen, PAWS matches the performance of the fully supervised model using the same features. This is shown in Fig.~\ref{fig:muscan}.  The fact that our model matches the dedicated fully supervised approach for $S/B\gtrsim 0.1\%$ also is a validation that the parameterized model is effective. PAWS achieves sensitivity to signals that are more than a factor of 10 weaker than for the classical weakly-supervised approach. This sensitivity is reaching the limit -- for a signal fraction of 0.03\%, the injected significance is only 0.13, and the fully supervised SIC is about 20. This means that the fully supervised approach would enhance the signal to only about 2. Therefore, no method can achieve useful performance below 0.03\%.
The lower panels of Fig.~\ref{fig:muscan} also demonstrate that the physics parameters are accurate.  Due to the subtle differences between the two- and three-prong models, we are not very sensitive to the exact value of $\alpha$.  

In order to show the robustness of our approach to the dimensionality of the feature space, we added 10 pure noise dimensions to the input~\cite{Finke:2023ltw,Freytsis:2023cjr} ($M=16$).  Each of these dimensions is sampled from a standard Normal distribution for both the signal and the background.  Since our pre-trained models only consider physical information, they effectively learn to ignore these additional features.  As a result, PAWS is unaffected by the noise.  In contrast, the traditional weakly supervised methods are strongly dependent on the noise and their sensitivity degrades by about a factor of 10 in $\mu$.

\section{Conclusions}
\label{sec:conclusions}

We have introduced a Prior-Assisted Weak Supervision (PAWS) approach to anomaly detection.   Like previous weakly-supervised methods, PAWS is based on training a classifier to distinguish data from a background-only reference dataset. We impose structure on the function class of classifiers through a pre-training step based on a set of signal models.   This trades off breadth for depth of sensitivity - we show that PAWS is able to improve the sensitivity of previous methods by a factor of 10--100 in signal cross section. While we require pre-specifying the signal class, it can (and in our case, does) contain a large number of models that would be infeasible to probe one at a time.  One could scan, in steps consistent with the experimental resolution, over the parameters of the models.  In addition to the computational challenge, this would also incur a large trial factor that PAWS does not pay since it uses pretraining and gradient descent.  Furthermore, the sensitivity of PAWS is at the limit of what is possible for any method using the same input features and so there is little to be lost compared to a dedicated approach.

There are a number of extensions of PAWS that would be exciting to explore in future work.  One direction would be to push the dimensionality of the parameter space.  In our numerical example, the parameter space was three-dimensional ($m_X, m_Y$, and $\alpha$) in addition to the signal fraction $\mu$.  Adding parameters would simply require adding features to the classifiers, but it may be advisable to not scan over them uniformly when building the parameterized classifier.  Another direction would be to push the dimensionality of the feature space.  In our numerical example, the feature space was six-dimensional.  Preliminary studies suggest that the setup works well when using the full phase space~\cite{Buhmann:2023acn}, but the performance is so good that more data is required for training and testing.  It would also be interesting to explore the sensitivity to signals that are similar to the ones used in the pre-training, but not exactly the same.  We found empirically that the model does not extrapolate well to $m_X$ and $m_Y$ outside of the training dataset, but this is not conclusive because the jet substructure also changes significantly (e.g. the decay products are no longer contained within a single jet).  While our example focused on resonant anomalies, we expect that the PAWS strategy would work in any scenario where data is compared to a background-only reference.  Combining PAWS with other weakly-supervised approaches for signal sensitivity and background estimation could lead to further improvements.  Ultimately, we expect that a variety of strategies with complementary assumptions and strengths will be required to achieve broad discovery potential at the LHC and beyond.

\section*{Data and code availability}

The original LHC Olympics datasets are available on Zenodo.  Our additional signal models, with the corresponding scripts to generate them, are also on Zenodo. The code we used to train and evaluate our models is available at \url{https://github.com/hep-lbdl/PAWS}.

\vspace{1cm}

\section*{Acknowledgments}

We thank V. Mikuni, R. Mastandrea, and D. Shih for helpful discussions and feedback on the manuscript.
CLC, GS, and BN are supported by the U.S. Department of Energy (DOE), Office of Science under contract DE-AC02-05CH11231.
This work was supported in part by the U.S. Department of Energy, Office of Science, Office of Workforce Development for Teachers and Scientists (WDTS) under the Science Undergraduate Laboratory Internships (SULI) program. 
This research used resources of the National Energy Research Scientific Computing Center, a DOE Office of Science User Facility supported by the Office of Science of the U.S. Department of Energy under Contract No. DE-AC02-05CH11231 using NERSC award HEP-ERCAP0021099.

\appendix

\bibliography{main}
\bibliographystyle{JHEP}

\end{document}